\documentstyle[12pt,aps]{revtex}
\newcommand{\bn}{\mbox{\boldmath $n$\unboldmath}}
\newcommand{\bl}{\mbox{\boldmath $l$\unboldmath}}
\newcommand{\bm}{\mbox{\boldmath $m$\unboldmath}}
\newcommand{\pr}{\frac{\partial}{\partial r}}
\newcommand{\pv}{\frac{\partial}{\partial v}} 
\newcommand{\pta}{\frac{\partial}{\partial \theta}}
\newcommand{\pvi}{\frac{\partial}{\partial \varphi}}

\newcommand{\spr}{\frac{\partial}{\partial r_*}}
\newcommand{\spv}{\frac{\partial}{\partial v_*}}
\newcommand{\spta}{\frac{\partial}{\partial \theta_*}}

\newcommand{\spdr}{\frac{\partial^2}{\partial r_*^2}}
\newcommand{\spdvr}{\frac{\partial^2}{\partial r_* \partial v_*}}
\newcommand{\spdra}{\frac{\partial^2}{\partial r_* \partial \theta_*}}
\newcommand{\spdri}{\frac{\partial^2}{\partial r_* \partial \varphi}}

\newcommand{\sta}{\sin\theta}
\newcommand{\cta}{\cos\theta}
\newcommand{\sda}{\sin^2\theta}
\newcommand{\cda}{\cos^2\theta}
\newcommand{\coa}{\cot\theta}
\newcommand{\sqd}{\sqrt{2}}
\newcommand{\cD}{\cal D}
\newcommand{\cL}{\cal L}
\newcommand{\cLd}{{\cal L}^{\dagger}}
\newcommand{\drH}{\dot{r}_H}
\newcommand{\ddrH}{\ddot{r}_H}
\newcommand{\prH}{r_H^{\prime}}
\newcommand{\ka}{\kappa}
\newcommand{\tkr}{2\kappa(r-r_H)}
\newcommand{\pprH}{r_H^{\prime\prime}}

\begin{document}
\pagestyle{myheadings}
\markboth{}{Wu and Cai}

\title{\Huge\bf Hawking Radiation of Dirac Particles \\ 
in a Variable-mass Kerr Space-time}
\author{S. Q. Wu\thanks{E-mail: sqwu@iopp.ccnu.edu.cn} and 
X. Cai\thanks{E-mail: xcai@ccnu.edu.cn} \\ 
\footnotesize \em
Institute of Particle Physics, Hua-Zhong 
Normal University, Wuhan 430079, China}
\maketitle

\vskip 1cm
\begin{abstract}
Hawking effect of Dirac particles in a variable-mass Kerr space-time is 
investigated by using a method called as the generalized tortoise coordinate 
transformation. The location and the temperature of the event horizon 
of the non-stationary Kerr black hole are derived. It is shown that the 
temperature and the shape of the event horizon depend not only on the time 
but also on the angle. However, the Fermi-Dirac spectrum displays 
a residual term which is absent from that of Bose-Einstein distribution.


\end{abstract}

\newpage
\baselineskip 20pt
\section{Introduction}

The fourth quarter of last century has witnessed various remarkable progress 
on several researches on black hole physics since Hawking's marked discovery 
\cite{Hawk}. One of these aspects is to reveal the thermal properties of
various black holes \cite{Ben}. Much efforts have been devoted to studying 
the thermal radiation of scalar fields or Dirac particles in the stationary 
axisymmetry black holes \cite{DR,WC,AM}. A popular method to determine the 
location and the temperature of the event horizon of an evaporating black 
hole is to calculate vacuum expectation value of the renormalized energy
momentum tensor \cite{HB}. But this method is very complicated, it gives 
only an approximate value of the location and that of the temperature. Thus
it is of limited use and meets great difficulties in many cases.

Zhao and Dai \cite{ZD} suggested a novel method of the generalized tortoise 
coordinate transformation (GTCT) which can give simultaneously the exact 
values both of the location and of the temperature of the event horizon of 
a non-stationary black hole. Basically, it is to reduce Klein-Gordon or Dirac 
equation in a black hole space-time to a standard wave equation near the 
event horizon by means of generalizing the common tortoise-type coordinate 
$r_* = r + \frac{1}{2\ka} \ln (r -r_H)$ in a static or stationary space-time 
\cite{San,DR,AM} ($\ka$ is the surface gravity of the studied event horizon) 
to a similar form in a non-static or non-stationary space-time \cite{ZD} by 
allowing that the location of the event horizon $r_H$ can be a function of 
the advanced time $v = t +r_*$ and/or of the angles $\theta, \varphi$. A key 
point in introducing the tortoise coordinate $r_*$ is that one must base upon 
the symmetric property of the considered space-time. For examples, the location 
of the event horizon is a constant ($r_H = 2M$) in the Schwarzschild black hole, 
while it is a function of the advanced time ($r_H = r_H(v)$) in a Vaidya-type 
space-time. The GTCT method has been applied to investigate thermal radiation 
of scalar particles in the case of non-stationary axisymmetric black holes 
\cite{KGNK}. Hawking effect of Dirac particles in the non-static black holes 
has also been studied successfully in \cite{DNS}. 

However, it is very difficult to discuss on the evaporation of Dirac 
particles in a non-stationary axisymmetric black hole. The difficulty 
lies in the non-separability of the radial and angular variables for 
Chandrasekhar-Dirac equation \cite{CP} in a non-stationary axisymmetry 
space-time. In this paper, we try to tackle with the thermal radiation 
of Dirac particles in a non-stationary Kerr space-time \cite{GHJW,CKC}. 
We consider the asymptotic behaviors of the first-order and second-order 
forms of Dirac equation near the event horizon. Using the relations between 
the first-order derivatives of Dirac spinorial components, we can eliminate 
the crossing-terms of the first-order derivatives in the second-order equation 
and recast each second-order equation to a standard wave equation near the 
event horizon. The location and the temperature of the event horizon are just 
the same as that obtained in the case of the thermal radiation of Klein-Gordon 
scalar field in a non-stationary Kerr space-time, but the Fermionic spectrum 
of Dirac particles displays another new effect dependent on the interaction 
between the spin of Dirac particles and the angular momentum of black holes.

Within the framework of Newman-Penrose formalism\cite{NP}, we derive in Sec.II 
the explicit form of Dirac equation in the variable-mass Kerr space-time. In
Sec.III, we introduce a GTCT according to the symmetry property of space-time
and investigate the asymptotic behavior of the first-order Dirac equation near 
the event horizon. Then the equation that determines the location of the event 
horizon are inferred from the vanishing determinant of the coefficients of the 
first-order derivative terms. A crucial step of our treatment in Sec. IV is to 
use the relations between the first-order derivative terms to eliminate the 
crossing term of the first derivatives in the second-order Dirac equation near 
the event horizon. Then we adjust the temperature parameter $\ka$ introduced 
in the GTCT so as to recast each second-order equation into a standard wave 
equation near the event horizon. In the meantime, we can get an exact 
expression of the Hawking temperature. Sec. V is devoted to separation of 
variables and to presenting the thermal radiation spectrum of Dirac particles 
according to the method of Damour-Ruffini-Sannan's \cite{DR,San}. Finally, we 
give a brief discussion about the spin-rotation effect. 

\section{Dirac Equation in a 
Non-stationary Kerr Space-time}

The variable mass Kerr solution \cite{GHJW,CKC} can be written in the 
advanced Eddington-Finkelstein system as
\begin{eqnarray}
&ds^2& = \frac{\Delta -a^2\sda}{\Sigma}dv^2 -2dvdr
+2a\sda dr d\varphi -\Sigma d\theta^2  \nonumber\\ 
&&+2\frac{r^2 +a^2 -\Delta}{\Sigma}a\sda dv d\varphi
-\frac{(r^2 +a^2)^2 -\Delta a^2\sda}{\Sigma}\sda d\varphi^2 \, , 
\end{eqnarray}
where $\Delta = r^2 -2M(v)r +a^2$, $\Sigma = r^2 +a^2\cda = \rho^*\rho$, 
$\rho^* = r +ia\cta$, $\rho = r -ia\cta$, and $v$ is standard advanced time. 
The mass $M$ depends on the advanced time $v$, but the specific angular
momentum $a$ is a constant.

We choose a complex null-tetrad $\{\bl, \bn, \bm, \overline{\bm}\}$ such 
that $\bl \cdot \bn = -\bm \cdot \overline{\bm} = 1$. Thus the covariant
one-forms can be written in the Kinnersley-like forms \cite{Kin} as
\begin{equation}
\begin{array}{ll}
\bl &= dv -a\sda d\varphi \, , \\ 
\bn &= \frac{\Delta}{2\Sigma} \left(dv -a\sda d\varphi \right) -dr \, , \\ 
\bm &= \frac{1}{\sqd\rho^*} \left\{i\sta \left[adv -(r^2 +a^2)d\varphi \right]
-\Sigma d\theta \right\} \, , \\ 
\overline{\bm} &= \frac{1}{\sqd\rho} \left\{-i\sta \left[adv -(r^2 +a^2)
d\varphi \right] -\Sigma d\theta \right\} \, ,  
\end{array}
\end{equation}
and their corresponding directional derivatives are
\begin{equation}
\begin{array}{ll}
D &= -\pr \, , \\ 
\underline{\Delta} &= \frac{r^2 +a^2}{\Sigma}\pv +\frac{\Delta}{2\Sigma}\pr
+\frac{a}{\Sigma}\pvi \, , \\ 
\delta &= \frac{1}{\sqd\rho^*} \left(ia\sta\pv +\pta 
+\frac{i}{\sta}\pvi \right) \, , \\ 
\overline{\delta} &= \frac{1}{\sqd\rho} \left(-ia\sta\pv +\pta
-\frac{i}{\sta}\pvi \right)\, .  
\end{array}
\end{equation}

Inserting for the following relations among spin-coefficients \cite{NP}
\begin{equation}
\begin{array}{ll}
&\epsilon -\tilde{\rho} = \frac{-r}{\Sigma}\, , \\
&\mu -\gamma = \frac{r -M}{2\Sigma} -\frac{ia\cta\Delta}{2\Sigma^2}\, ,\\
&\tilde{\pi} -\alpha = \frac{\coa}{2\sqd\rho} 
-\frac{ira\sta}{\sqd\Sigma\rho}\, , \\
&\beta -\tau = \frac{\coa}{2\sqd\rho^*}
-\frac{a^2\sta\cta}{\sqd\Sigma\rho^*}\, ,
\end{array}
\end{equation}
into four coupled Chandrasekhar equations \cite{CP} in the Newman-Penrose 
formalism \cite{NP}  
\begin{equation}
\begin{array}{ll}
&(D +\epsilon -\tilde{\rho})F_1 +(\overline{\delta} +\tilde{\pi}
-\alpha)F_2 = \frac{i\mu_0}{\sqd}G_1 \, , \\
&(\underline{\Delta} +\mu -\gamma )F_2
+(\delta +\beta -\tau)F_1 = \frac{i\mu_0}{\sqd}G_2 \, ,\\
&(D +\epsilon^* -\tilde{\rho}^*)G_2 -(\delta +\tilde{\pi}^*
-\alpha^*)G_1 = \frac{i\mu_0}{\sqd}F_2 \, , \\
&(\underline{\Delta} +\mu^* -\gamma^*)G_1 -(\overline{\delta}
+\beta^* -\tau^*)G_2 = \frac{i\mu_0}{\sqd}F_1 \, , 
\end{array}
\end{equation}
where $\mu_0$ is the mass of Dirac particles, one can obtain 
\begin{equation}
\begin{array}{rr}
-\left(\pr +\frac{r}{\Sigma} \right)F_1 
+\frac{1}{\sqd\rho} \left({\cL} -\frac{ira\sta}{\Sigma} \right)F_2 
=& \frac{i\mu_0}{\sqd}G_1 \, , \\
\frac{\Delta}{2\Sigma} \left({\cD} -\frac{ia\cta}{\Sigma} \right)F_2 
+\frac{1}{\sqd\rho^*} \left({\cLd} -\frac{a^2\sta\cta}{\Sigma} \right)F_1
=& \frac{i\mu_0}{\sqd}G_2 \, , \\
-\left(\pr +\frac{r}{\Sigma} \right)G_2 
-\frac{1}{\sqd\rho^*} \left({\cLd} +\frac{ira\sta}{\Sigma} \right)G_1 
=& \frac{i\mu_0}{\sqd}F_2 \, , \\
\frac{\Delta}{2\Sigma} \left({\cD} +\frac{ia\cta}{\Sigma} \right)G_1 
-\frac{1}{\sqd\rho} \left({\cL} -\frac{a^2\sta\cta}{\Sigma} \right)G_2
=& \frac{i\mu_0}{\sqd}F_1 \, , \label{DCP}
\end{array}
\end{equation}
here we have defined operators
\begin{eqnarray*}
&{\cD}& = \pr +\Delta^{-1} \left[r-M +2a\pvi +2(r^2 +a^2)\pv \right] \, , \\ 
&{\cL}& = \pta +\frac{1}{2}\coa -\frac{i}{\sta}\pvi -ia\sta\pv \, , \\
&{\cLd}& = \pta +\frac{1}{2}\coa +\frac{i}{\sta}\pvi +ia\sta\pv \, .
\end{eqnarray*}

By substituting
$$F_1=\frac{1}{\sqrt{2\Sigma}}P_1 \, ,
~~~F_2=\frac{\rho}{\sqrt{\Sigma}}P_2 \, ,
~~~G_1=\frac{\rho^*}{\sqrt{\Sigma}}Q_1 \, ,
~~~G_2=\frac{1}{\sqrt{2\Sigma}}Q_2 \, , $$
into Eq. (\ref{DCP}), they have the form
\begin{equation}
\begin{array}{ll}
-\pr P_1 +{\cL} P_2 = i\mu_0\rho^* Q_1 \, , 
&~~~~\Delta{\cD} P_2 + {\cLd} P_1 = i\mu_0 \rho^* Q_2 \, ,\\
-\pr Q_2 -{\cLd} Q_1 = i\mu_0\rho P_2 \, , 
&~~~~\Delta{\cD} Q_1 - {\cL} Q_2 = i\mu_0\rho P_1 \, . \label{reDP}
\end{array}
\end{equation}

\section{Event Horizon}
 
Eq. (\ref{reDP}) can not be decoupled except in the case of Kerr black hole 
\cite {CP} ($M = const$) or in the case of Vaidya space-time \cite{BV} ($a=0$). 
However, to deal with the problem of Hawking radiation, one may concern about 
the behavior of Eq. (\ref{reDP}) near the horizon only. As the space-time we 
consider at present is symmetric about the $\varphi$-axis, one can introduce 
a GTCT as does in Ref. \cite{KGNK} 
\begin{equation}
\begin{array}{ll}
&r_* = r +\frac{1}{2\ka} \ln [r -r_H(v, \theta)] \, ,\\
&v_* = v -v_0 \, , ~~~~~~~~~~~~\theta_* = \theta -\theta_0 \, , \label{trans}
\end{array}
\end{equation}
where $r_H = r_H(v, \theta)$ is the location of the event horizon, and $\ka$ 
is an adjustable parameter. All parameters $\ka$, $v_0$ and $\theta_0$ are 
constant under the tortoise transformation. The derivatives change 
correspondingly as
\begin{eqnarray*}
\begin{array}{ccc}
\pr &= \spr +\frac{1}{\tkr}\spr \, ,  \\
\pv &= \spv -\frac{\drH}{\tkr}\spr \, , \\
\pta &= \spta -\frac{\prH}{\tkr}\spr  \, .
\end{array}
\end{eqnarray*}
where $\drH=\frac{\partial r_H}{\partial v}$ is the rate of the event
horizon varying in time, $\prH=\frac{\partial r_H}{\partial \theta}$ is 
its rate changing with the angle $\theta$.

Under the transformations (\ref{trans}), Eq. (\ref{reDP}) can be reduced to 
\begin{equation}
\begin{array}{ll}
&\spr P_1 +\left(\prH -ia\sta_0 \drH \right)\spr P_2 = 0 \, , \\ 
&-\left(\prH +ia\sta_0 \drH \right)\spr P_1 +\left[\Delta_H -2(r_H^2 +a^2) 
\drH \right]\spr P_2 = 0 \, , \label{trDPP} 
\end{array}
\end{equation}
and
\begin{equation}
\begin{array}{ll}
&\left(\prH +ia\sta_0 \drH \right)\spr Q_1 -\spr Q_2 = 0 \, , \\ 
&\left[\Delta_H -2(r_H^2 +a^2) \drH \right]\spr Q_1 
+\left(\prH -ia\sta_0 \drH \right)\spr Q_2 = 0 \, , \label{trDPQ}
\end{array}
\end{equation}
after being taken the $r \rightarrow r_H(v_0, \theta_0)$, $v \rightarrow 
v_0$ and $\theta \rightarrow \theta_0$ limits. We have denoted $\Delta_H 
= r_H^2 -2M(v)r_H +a^2 $.

If the derivatives $\spr P_1$, $\spr P_2$, $\spr Q_1$ and $\spr Q_2$ in Eqs. 
(\ref{trDPP},\ref{trDPQ}) are nonzero, the existence condition of non-trial 
solutions for $P_1$, $P_2$, $Q_1$ and $Q_2$ is that the determinant of Eqs. 
(\ref{trDPP},\ref{trDPQ}) vanishes, which gives the following equation to 
determine the location of horizon  
\begin{equation}
\Delta_H -2(r_H^2 +a^2)\drH +a^2\sda_0 {\drH}^2 +{\prH}^2 = 0 \, ,
\label{loeh}
\end{equation}
Eq. (\ref{loeh}) just is the equation which can be inferred from the 
null-surface condition $g^{ij}\partial_i f \partial_j f = 0$, and 
$f(v,r,\theta) = 0$, namely $r=r(v, \theta)$. 

The location of the event horizon is in accord with that obtained in the case 
of Klein-Gordon field equation \cite{KGNK}. It means that the location of the 
horizon is shown as
\begin{equation}
r_H = \frac{M \pm [M^2 -(a^2\sda_0 {\drH}^2 +{\prH}^2)(1 -2\drH) 
-a^2(1 -2\drH)^2]^{1/2}}{1 -2\drH} \, . \label{loca}
\end{equation}

\section{Hawking Temperature}

Apparently one can find that the Chandrasekhar-Dirac equations (\ref{reDP}) 
could be satisfied by identifying $Q_1$, $Q_2$ with $P_2^*$, $-P_1^*$, 
respectively. So he may deal with a pair of components $P_1$, $P_2$ only. 
Now, we rewrite the Dirac equation in a two-component form as
\begin{equation}
\left(\begin{array}{cc}
-\pr & {\cL} \\
{\cLd} & \Delta {\cD} 
\end{array} \right) 
\left(\begin{array}{c}
P_1 \\ P_2
\end{array}\right) = i\mu_0 \rho^*
\left(\begin{array}{c}
Q_1 \\ Q_2
\end{array}\right) \, , ~~~~
\left(\begin{array}{cc}
-\Delta {\cD} & {\cL} \\
{\cLd} & \pr
\end{array}\right)
\left(\begin{array}{c}
Q_1 \\ Q_2
\end{array}\right) = -i\mu_0 \rho
\left(\begin{array}{c}
P_1 \\ P_2
\end{array}\right) \, ,
\end{equation}
and obtain its corresponding second-order form for the two-component 
spinor ($P_1, P_2$) as follows 
\begin{equation}
\begin{array}{ll}
&\left[\Delta {\cD}\pr +{\cL}{\cLd} +\frac{1}{\rho^*}
\left(ia\sta {\cLd} -\Delta\pr \right) \right] P_1   \\
&~~~~~~~~~~~~~~~~~~+\left\{\frac{\Delta}{\rho^*} \left({\cL} 
+ia\sta {\cD} \right) +ia\sta \left[2\dot{M}r\pr +\dot{M} \right] 
\right\} P_2 = \mu_0^2 \Sigma P_1 \, ,  \\
&\left[\pr\Delta {\cD} +{\cLd}{\cL} +\frac{1}{\rho^*} \left(ia\sta {\cL}
-\Delta {\cD} \right) \right] P_2 -\frac{1}{\rho^*} \left({\cLd} 
+ia\sta\pr \right) P_1 = \mu_0^2 \Sigma P_2 \, , \label{socd}
\end{array}
\end{equation}

Given the GTCT in Eq. (\ref{trans}) and after some tedious calculations, the 
limiting form of Eq. (\ref{socd}), when $r$ approaches $r_H(v_0, \theta_0)$, 
$v$ goes to $v_0$ and $\theta$ goes to $\theta_0$, reads
\begin{eqnarray}
&&\left[\frac{r_H(1 -2\drH) -M}{\ka} +2\Delta_H -2\drH (r_H^2 +a^2) 
\right]\spdr P_1 +2a(1 -\drH)\spdri P_1  \nonumber \\
&&+2(r_H^2 +a^2 -\drH a^2\sda_0)\spdvr P_1 -2\prH\spdra P_1
-\{r_H(1 -3\drH) -M  \nonumber \\ 
&&+\pprH +\ddrH a^2\sda_0 +\prH\coa_0 +\frac{\Delta_H -\drH (r_H^2 +a^2) 
+ia\sta_0 \prH}{\rho_H^*} \}\spr P_1 \nonumber \\ 
&&+\left\{2i\dot{M}r_H a\sta_0 -\frac{\Delta_H [\prH -ia\sta_0 (\drH  +1)] 
+2ia\sta_0 \drH (r_H^2 +a^2)}{\rho_H^*} \right\}\spr P_2 = 0 \, , \label{wone}
\end{eqnarray}
and
\begin{eqnarray}
&&\left[\frac{r_H(1 -2\drH) -M}{\ka} +2\Delta_H -2\drH (r_H^2 +a^2)
\right]\spdr P_2 +2(r_H^2 +a^2 -\drH a^2\sda_0)\spdvr P_2 \nonumber \\
&&+2a(1 -\drH)\spdri P_2 -2\prH\spdra P_2 -\{M -r_H(1 +\drH) 
+\pprH +\ddrH a^2\sda_0 +\prH \coa_0 \nonumber \\ 
&&+\frac{\Delta_H -\drH (r_H^2 +a^2) +ia\sta_0 \prH}{\rho_H^*} \}\spr P_2 
+\frac{\prH +ia\sta_0 (\drH -1)}{\rho_H^*}\spr P_1 = 0 \, . \label{wtwo} 
\end{eqnarray}
where $\rho_H^* =r_H +ia\cta_0$. In the calculations, we have used the
L'H\^osptial's rule to treat an infinite form of $0 \over 0$-type.

One can adjust the parameter $\ka$ such that it satisfies
\begin{equation}
\frac{r_H(1 -2\drH) -M}{\ka} +2\Delta_H -2\drH (r_H^2 +a^2)
= r_H^2 +a^2 -\drH a^2\sda_0 \, ,
\end{equation}
which means the temperature of the horizon is
\begin{eqnarray}
\ka &=& \frac{r_H(1 -2\drH) -M}{r_H^2 +a^2 -\drH a^2\sda_0 
-2\Delta_H +2\drH (r_H^2 +a^2)} \nonumber \\
&=& \frac{r_H(1 -2\drH) -M}{(r_H^2 +a^2 -\drH 
a^2\sda_0)(1 -2\drH) +2{\prH}^2} \, , \label{temp}
\end{eqnarray}
where we have used Eq. (\ref{loeh}).

Using Eq. (\ref{trDPP}), namely
$$\spr P_1 = -\left(\prH -ia\sta_0 \drH \right)\spr P_2 \, ,~~~~
\spr P_2 = \frac{\prH +ia\sta_0 \drH}{\Delta_H -2(r_H^2 +a^2)}\spr P_1  
\, , $$
Eq. (\ref{wone}) and Eq. (\ref{wtwo}) can be recast into the standard
wave equation near the horizon in an united form
\begin{eqnarray}
\spdr \Psi +2\spdvr \Psi +2\Omega \spdri \Psi +2C_3 \spdra \Psi 
+2(C_2 +iC_1) \spr \Psi = 0 \, , \label{wave}
\end{eqnarray}
where $\Omega$ is the angular velocity of the event horizon of the 
evaporating Kerr black hole,
$$\Omega = \frac{a(1 -\drH)}{r_H^2 +a^2 -\drH a^2\sda_0} \, , 
~~~~ C_3 = \frac{-\prH}{r_H^2 +a^2 -\drH a^2\sda_0} \, ,$$
while both $C_1$ and $C_2$ are real constants, 
\begin{eqnarray*}
C_2 &=& \frac{-1}{2(r_H^2 +a^2 -\drH a^2\sda_0)} 
\left[r_H(1 -4\drH) -M +\pprH  \right. \\
&&\left.+\ddrH a^2\sda_0 +\prH \coa_0 +\frac{2\dot{M}r_H\drH 
a^2\sda_0}{\Delta_H -2\drH (r_H^2 +a^2)} \right] \, ,  \\
C_1 &=& \frac{1}{2(r_H^2 +a^2 -\drH a^2\sda_0)} 
\left[-\drH a\cta_0 +\frac{2\dot{M}r_H\prH 
a\sta_0}{\Delta_H -2\drH (r_H^2 +a^2)} \right] \, ,  \\
\end{eqnarray*}
for $\Psi = P_1$, and
\begin{eqnarray*}
C_2 &=& -\frac{M -r_H +\pprH +\ddrH a^2\sda_0 
+\prH \coa_0}{2(r_H^2 +a^2 -\drH a^2\sda_0)} \, ,  \\
C_1 &=& \frac{\drH a\cta_0}{2(r_H^2 +a^2 -\drH a^2\sda_0)} \, ,
\end{eqnarray*}
for $\Psi = P_2$.

\section{Thermal Radiation Spectrum}

Now separating variables as $\Psi = R(r_*)\Theta(\theta_*)
e^{i(m\varphi -\omega v_*)}$ , one has
\begin{equation}
\begin{array}{ll}
&\Theta^{\prime} = \lambda \Theta \, ,  \\
&R^{\prime\prime} +2(C_0 +iC_1 +im\Omega -i\omega)R^{\prime} = 0 \, , 
\end{array}
\end{equation}
where $\lambda$ is a real constant introduced in the separation of 
variables, $C_0 = \lambda C_3 +C_2$. The solutions are
\begin{equation}
\begin{array}{ll}
&\Theta = e^{\lambda \theta_*} \, ,  \\
& R \sim e^{2i(\omega -m\Omega -C_1)r_* -2C_0r_*} \, ; R_0 \, . 
\end{array}
\end{equation}

The ingoing wave solution and the outgoing wave solution 
to Eq. (\ref{wave}) are, respectively,
\begin{equation}
\begin{array}{ll}
&\Psi_{\rm in} = e^{-i\omega v_* +im\varphi +\lambda \theta_*} \, , \\
&\Psi_{\rm out} = e^{-i\omega v_* +im\varphi +\lambda \theta_*}
e^{2i(\omega -m\Omega -C_1)r_* -2C_0r_*}\, ,~~~~ (r > r_H) \, . 
\end{array}
\end{equation}

The outgoing wave $\Psi_{\rm out}$ is not analytic at the event horizon
$r = r_H$, but can be analytically continued from the outside of the hole 
into the inside of the hole by the lower complex $r$-plane,
$$ (r -r_H) \rightarrow (r_H -r) e^{-i\pi}$$
to 
\begin{equation}
\tilde{\Psi}_{\rm out} = e^{-i\omega v_* +im\varphi +\lambda \theta_*}
e^{2i(\omega -m\Omega -C_1)r_* -2C_0r_*}e^{i\pi C_0/\ka}
e^{\pi(\omega -m\Omega -C_1)/\ka} \, , ~~~~ (r < r_H) \, . 
\end{equation}

The relative scattering probability at the event horizon is
\begin{equation}
\left|\frac{{\Psi}_{\rm out}}{\tilde{\Psi}_{\rm out}}\right|^2
= e^{-2\pi(\omega -m\Omega -C_1)/\ka} \, . 
\end{equation}
Following the method suggested by Damour and Ruffini \cite{DR}, and extended 
by Sannan \cite{San}, the Fermionic spectrum of Hawking radiation of Dirac 
particles from the black hole is easily obtained
\begin{equation} 
\langle {\cal N}(\omega) \rangle = 
\frac{1}{e^{(\omega -m\Omega -C_1)/T_H } +1} \, , ~~~~~~~~~
T_H = \frac{\ka}{2\pi} \, . \label{sptr}
\end{equation} 

\section{Conclusion}

Equations (\ref{loca}) and (\ref{temp}) give the location and the temperature 
of event horizon of the variable-mass Kerr black hole, which depend not only 
on the advanced time $v$ but also on the angle $\theta$. They can recover the 
well-known results previously obtained by others. Eq. (\ref{sptr}) shows the 
thermal radiation spectrum of Dirac particles in the non-stationary Kerr 
space-time, in which a residual term $C_1$ appears. The difference between
Bosonic spectrum and Fermionic spectrum is that $C_1$ is absent in the spectrum 
of Klein-Gordon scalar particles. Also $C_1$ vanishes when a black hole is 
stationary ($\drH = \prH = 0$) or it has a zero angular momentum ($a = 0$). 
Thus this new effect maybe arise from the interaction between the spin of 
Dirac particle and the evaporating black hole. 

In conclusion, we succeed in dealing with Chandrasekhar-Dirac equation in 
the non-stationary Kerr black hole. Under the generalized tortoise coordinate
transformation, each second-order equation induced from of Chandrasekhar-Dirac 
equation takes the standard form of wave equation near the event horizon, to 
which separation of variables is possible. A key step in our arguments is that 
we have considered the asymptotic form of the first-order equation near the
event horizon, and obtained formulas relating the first-order derivatives of
different component which enable us to eliminate the crossing-terms in the
second-order equation. The location and the temperature of event horizon of 
the variable-mass Kerr black hole are just the same as that obtained in the 
discussing on thermal radiation of Klein-Gordon particles in the same 
space-time. But the spectrum of Dirac particles displays another new effect 
which is absent from the spectrum of Klein-Gordon particles. This effect
is a kind of spin-rotation coupling effect.

\section*{Appendix: Newman-Penrose coefficients}
\setcounter{equation}{0}
\renewcommand{\theequation}{A\arabic{equation}}

In this appendix, we make use of the method of exterior differentiation 
to calculate the Newman-Penrose coefficients in the non-stationary Kerr
space-time. The complex null-tetrad $\{\bl, \bn, \bm, \overline{\bm}\}$ 
satisfies the orthogonal conditions $\bl \cdot \bn = -\bm \cdot \overline{\bm} 
= 1$. We choose the covariant one-forms similar to the Kinnersley forms as 
\begin{equation}
\begin{array}{ll}
\bl &= dv -a\sda d\varphi \, , \\ 
\bn &= \frac{\Delta}{2\Sigma} \left(dv -a\sda d\varphi \right) -dr \, ,\\ 
\bm &= \frac{1}{\sqd\rho^*} \left\{i\sta \left[adv -(r^2 +a^2)d\varphi 
\right] -\Sigma d\theta \right\} \, , \\ 
\overline{\bm} &= \frac{1}{\sqd\rho} \left\{-i\sta \left[adv -(r^2 +a^2)
d\varphi \right] -\Sigma d\theta \right\} \, . 
\end{array}
\end{equation}

The exterior differentiation of the one-forms bases gives
\begin{equation}
\begin{array}{ll}
d\bl &= \frac{\sqd a^2\sta\cta}{\Sigma} \left(\frac{\bm}{\rho}
+\frac{\overline{\bm}}{\rho^*} \right) \wedge \bl 
-\frac{2ia\cta}{\Sigma} \bm \wedge \overline{\bm} \, , \\ 
d\bn &= \left(\frac{r\Delta}{\Sigma^2} -\frac{r -M}{2\Sigma} \right) \bn 
\wedge \bl -\frac{i\dot{M}ra\sta\cta}{2\sqd \Sigma} \left(\frac{\bm}{\rho}
-\frac{\overline{\bm}}{\rho^*} \right) \wedge \bl 
-\frac{ia\cta\Delta}{\Sigma^2} \bm \wedge \overline{\bm} \, ,\\
d\bm &= \frac{\sqd ira\sta}{\Sigma\rho^*} \bn \wedge \bl -\frac{1}{\rho}
\left(\bn -\frac{\Delta}{2\Sigma} \bl \right) \wedge \bm +\frac{1}{\sqd\rho^*}
\left(\coa +\frac{ia\sta}{\rho^*} \right) \bm \wedge \overline{\bm} \, , \\ 
d\overline{\bm} &= -\frac{\sqd ira\sta}{\Sigma\rho} \bn \wedge \bl 
-\frac{1}{\rho^*} \left(\bn -\frac{\Delta}{2\Sigma} \bl \right) \wedge
\overline{\bm} -\frac{1}{\sqd\rho} \left(\coa -\frac{ia\sta}{\rho}
\right) \bm \wedge \overline{\bm} \, .  
\end{array}
\end{equation}

In general, the common relations among the exterior differentiations of 
one-forms bases hold:
\begin{equation}
\begin{array}{ll}
d\bl &= (\epsilon +\epsilon^*) \bn \wedge \bl 
+(\alpha +\beta^* -\tau^*) \bl \wedge \bm 
+(\alpha^* +\beta -\tau) \bl \wedge \overline{\bm}  \\ 
&~~~~ -\tilde{\kappa}^* \bn \wedge \bm 
-\tilde{\kappa} \bn \wedge \overline{\bm}  
+(\tilde{\rho} -\tilde{\rho}^*) \bm \wedge \overline{\bm} \, , \\ 
d\bn &= (\gamma +\gamma^*) \bn \wedge \bl +\nu \bl \wedge \bm 
+\nu^* \bl \wedge \overline{\bm}  \\ 
&~~~~ +(\tilde{\pi} -\alpha -\beta^*) \bn \wedge \bm 
+(\tilde{\pi}^* -\alpha^* -\beta) \bn \wedge \overline{\bm}  
+(\mu -\mu^*) \bm \wedge \overline{\bm} \, ,\\
d\bm &= (\tilde{\pi}^* +\tau) \bn \wedge \bl 
+(\gamma -\gamma^* +\mu^*) \bl \wedge \bm 
+\tilde{\lambda}^* \bl \wedge \overline{\bm}  \\ 
&~~~~ +(\epsilon -\epsilon^* -\tilde{\rho}) \bn \wedge \bm 
-\sigma \bn \wedge \overline{\bm}  
+(\beta -\alpha^*) \bm \wedge \overline{\bm} \, , \\ 
d\overline{\bm} &= (\tilde{\pi} +\tau^*) \bn \wedge \bl 
+\tilde{\lambda} \bl \wedge \bm 
+(\gamma^* -\gamma +\mu) \bl \wedge \overline{\bm}  \\ 
&~~~~ -\sigma^* \bn \wedge \bm 
+(\epsilon^* -\epsilon -\tilde{\rho}^*) \bn \wedge \overline{\bm}  
+(\alpha -\beta^*) \bm \wedge \overline{\bm} \, .  
\end{array}
\end{equation}
where a star $*$ stands for complex conjugate as usual.

It is not difficult to determine the twelve Newman-Penrose complex coefficients 
\cite{NP} in the above null-tetrad as follows 
\begin{eqnarray}
\begin{array}{lll}
\tilde{\kappa} = 0\, , &\tilde{\lambda} = 0\, , &\sigma = 0\, , \\
\tilde{\rho} = \frac{1}{\rho^*}\, ,&\epsilon = -\frac{ia\cta}{\Sigma}\, ,
&\gamma = \frac{r\Delta}{2\Sigma^2} -\frac{r -M}{2\Sigma}\, , \\
\mu = \frac{\Delta}{2\Sigma\rho^*}\, , &\tau = \frac{ia\sta}{\sqd\rho^{*2}}\, ,
&\nu = \frac{\dot{M}ria\sta}{\sqd\Sigma\rho}\, , \\
\alpha = \tilde{\pi} -\beta^*\, , &\tilde{\pi} = -\frac{ia\sta}{\sqd\Sigma}\, ,
&\beta = \frac{\coa}{2\sqd\rho^*} +\frac{ira\sta}{\sqd\Sigma\rho^*}\, .
\end{array}
\end{eqnarray}

\vskip 0.5cm
\noindent
{\bf Acknowledgment}

S.Q. Wu is very grateful to Dr. Jeff Zhao at Motomola Company for his
long-term helps. Thanks is also devoted to Prof. E. Takasugi at Osaka
University for having sent their preprints. This work is supported in
part by the NSFC in China.

\end{document}